# Engineering of spin-lattice relaxation dynamics by digital growth of diluted magnetic semiconductor CdMnTe

M. K. Kneip[1], D. R. Yakovlev[1,2], M. Bayer[1], G. Karczewski[3], T. Wojtowicz[3], J. Kossut[3]

[1]*Experimentelle Physik II, Universität Dortmund, D-44221 Dortmund, Germany*
[2]*A.F. Ioffe Physico-Technical Institute, Russian Academy of Sciences, 194017 St. Petersburg, Russia*
[3]*Institute of Physics, Polish Academy of Sciences, PL-02668 Warsaw, Poland*

The technological concept of "digital alloying" offered by molecular-beam epitaxy is demonstrated to be a very effective tool for tailoring static and dynamic magnetic properties of diluted magnetic semiconductors. Compared to common "disordered alloys" with the same Mn concentration, the spin-lattice relaxation dynamics of magnetic Mn ions has been accelerated by an order of magnitude in (Cd,Mn)Te digital alloys, without any noticeable change in the giant Zeeman spin splitting of excitonic states, i.e. without effect on the static magnetization. The strong sensitivity of the magnetization dynamics to clustering of the Mn ions opens a new degree of freedom for spin engineering.

PACS: 75.50.Pp,  78.55.Et,  78.20.Ls,  85.75.-d

II-VI diluted magnetic semiconductors (DMS) like (Cd,Mn)Te or (Zn,Mn)Se are well-known model materials for testing spintronic concepts [1]. A variety of magneto-optical and magneto-transport effects originates from the strong exchange interaction of free carriers with localized magnetic moments of $Mn^{2+}$ ions [2]. The magnetic properties of the Mn ion system, which depend strongly on the Mn concentration, play a key role in these effects. For example, neighbouring Mn ions interact antiferromagnetically, which leads to the formation of high ordered clusters and spin-glass phases at higher Mn concentrations. Also the dynamical properties of the localized Mn spins, namely spin decoherence and spin-lattice relaxation (SLR), are controlled by concentration dependent exchange interactions between the Mn ions [3]. The SLR time varies by several orders of magnitude from milliseconds down to nanoseconds with increasing Mn content [3, 4].

For the common "disordered alloys" with a random distribution of magnetic ions in the cation sublattice, static and dynamic magnetic properties, which are controlled by the Mn ions, are strongly correlated with each other, as both depend on Mn concentration. E.g., by designing a specific value for the giant Zeeman splitting of carriers, which is proportional to the static magnetization, we also determine the SLR time for the Mn system, which corresponds to the chosen Mn content. A way to overcome this limitation is offered by the technological concept of "digital alloying", because static and dynamic properties of DMS are governed by different mechanisms. Paramagnetic Mn spins give the main contribution to the static magnetization, and their coupling into antiferromagnetic Mn-Mn clusters for increasing Mn content is unfavourable in this respect. On the other hand, the SLR dynamics is controlled by anisotropic exchange interactions of Mn ions in the clusters. We will show in this Letter that digital alloying of DMS provides considerable changes in clustering which, in turn, modifies the magnetization dynamics, while keeping about constant the number of paramagnetic spins, which controls the static magnetization.

The concept of digital growth technique [5, 6] has been brought about by atomic precision of molecular beam epitaxy (MBE). During the growth some constituent of the structure, say AlAs, with a certain thickness (frequently sub-monolayer) is introduced into a base material, say GaAs, at strictly predefined positions. For example, this technique allows for engineering almost at will the shape of the confining potential in quantum structures in a digital fashion. If the constituents are built-in periodically with a very small period of the order of a single monolayer, the result is a "digital alloy" (DA) [7]. This is to be contrasted with the disordered alloys in which both constituents are introduced simultaneously to form mixed crystals (e.g. $Ga_{1-x}Al_xAs$). In the area of DMS the concept of digital magnetic quantum wells was introduced in 1995 [8, 9]. The possibility of a unique spin splitting and spin dynamics engineering was demonstrated for MnSe/(Zn,Cd)Se digital alloys [9] and for (Cd,Mn)Te/CdTe digitally-graded magnetic quantum well structures [10]. A strong variation of the carrier spin dynamics with the DA parameters has been found. However, no dependence of the Mn spin dynamics has been established yet. In this Letter we show that the spin dynamics of magnetic Mn ions can be engineered to a desired value by a proper choose of DA parameters.

All samples were grown by MBE on (100)-oriented GaAs substrates after depositing a 4.2-µm-thick $Cd_{0.8}Mg_{0.2}Te$ buffer layer, which compensates the lattice constant mismatch between GaAs substrate and II-VI heterostructure. The three studied samples contain 23-nm-thick layers [corresponding to about 70 monolayers (ML)] of a $Cd_{0.95}Mn_{0.05}Te$/CdTe digital alloy confined from both sides by $Cd_{0.8}Mg_{0.2}Te$ barriers, to form digital magnetic quantum wells [8]. The structures are labelled according to the thickness of magnetic and nonmagnetic sheets given in the monolayers: the first sample 1×3DA consists of 18 periods of 1ML/3ML, the second sample 2×6DA has 9 periods of 2ML/6ML and the third sample 3×9DA has 6 periods of 3ML/9ML (see Fig. 1). This choice of layer thicknesses provides the same average Mn concentration $x_{DA} \approx 0.013$ in the three DA samples. It is known for DMS heterostructures that the Mn ions at the interface can diffuse into nonmagnetic layers over a depth of 1-2 ML [11, 12]. Therefore, in DA samples the real Mn profile may differ from the nominal technological design. Additionally two reference disordered alloy samples with 1-µm-thick layers of $Cd_{0.985}Mn_{0.015}Te$ and $Cd_{0.96}Mn_{0.04}Te$ were grown for comparative studies.

The SLR dynamics was measured by an all-optical technique. The Mn spin system, polarized by an external magnetic field, was heated by a pulsed laser and the dynamical shift of the photoluminescence line was detected (for details see Ref. [4]). Photoexcitation was provided by 7-ns laser pulses at 355 nm wavelength with the power of 1.2 mW and the repetition rate of 3 kHz. Instead of using a *cw* laser as in previous studies for tracing photoluminescence at delays exceeding the duration of the heating pulse, we use a modulated semiconductor laser (375 nm, power density 72.5 mW/cm$^2$ at 3 kHz repetition rate and 250 ns pulse length). The pulsed operation mode of this laser allows us to reduce drastically (by two or three orders of magnitude) the contribution of this laser's illumination to the unwanted background heating of the Mn system, which is considerable for low Mn concentrations. Photoluminescence spectra were detected with a gated charge-coupled-device camera (time resolution of 5 ns) combined with a 0.5-m spectrometer. Experiments were performed at the temperature $T = 1.7$ K with samples immersed in pumped liquid helium. External magnetic fields $B$ up to 10 T were applied parallel to the structure growth axis (Faraday geometry).

Typical PL spectra of DA samples measured under *cw* excitation are given in the inset of Fig. 2 (a). Narrow linewidths, not exceeding 4 meV, evidence the high structural and optical quality of the digital alloys. In external magnetic field the PL lines shift to lower energies [shown by symbols in Fig. 2 (a)]. This shift is equal to one half of the giant Zeeman splitting of heavy-hole excitons $\Delta E_Z$ [2]. It is proportional to the magnetization of the Mn spin system $M(B)$, which gives an optical access to the static magnetization.

$$\Delta E_Z(B,T) = \frac{(\beta - \alpha)}{\mu_B g_{Mn}} M(B,T) \qquad (1)$$

$$M(B,T) = \mu_B g_{Mn} x N_0 S_{eff}(x) \, \mathrm{B}_{5/2}\left[\frac{5 g_{Mn} \mu_B B}{2 k_B (T_{Mn} + T_0(x))}\right] \qquad (2)$$

Here $N_0\alpha = 0.22$ eV and $N_0\beta = -0.88$ eV are the exchange constants in Cd$_{1-x}$Mn$_x$Te for the conduction and valence band, respectively [2]. $N_0$ is the inverse unit-cell volume and $x$ is the Mn mole fraction. B$_{5/2}$ is the modified Brillouin function; $T_{Mn}$ is the Mn spin temperature; $g_{Mn} = 2$ is the g-factor of the Mn$^{2+}$ ions. $S_{eff}$ is the effective spin and $T_0$ is the effective temperature. These parameters permit a phenomenological description of the antiferromagnetic Mn-Mn exchange interaction.

All three DA samples show very similar Zeeman shifts. The Mn contents $x_{DA}$ extracted from fits of the shifts by Eqs. (1) and (2) are 0.0147, 0.0130 and 0.0123 for the 1×3DA, 2×6DA and 3×9DA samples, respectively. These values are very close to the average Mn concentration in our digital alloys $x_{DA} \approx 0.013$, if the distribution of Mn component were assumed to be random. From these data we conclude that the static magnetic properties of the studied DA are weakly affected by the DA growth parameters. Small reduction of the Zeeman shift for the DA with thicker Cd$_{0.95}$Mn$_{0.05}$Te layers can be explained by binding the paramagnetic Mn spins into clusters.

The dynamics of spin-lattice relaxation of Mn ions has been measured in a magnetic field of 3 T from the time-resolved energy shift of the PL maxima $\Delta E_{PL}(t)$ induced by the laser pulses. In Fig. 2 (b) these data are given normalized to the maximum shift $\Delta E_{PL}^{max}$. Closed symbols show data for the three DA samples and open symbols are data for the reference disordered alloys. The SLR times $\tau_{SLR}$ of 27 and 2.2 μs measured for the disordered alloys with $x=0.015$ and 0.04, respectively, are in good agreement with literature data (see Refs. [4, 13] and References therein). It is remarkable that $\tau_{SLR}$ for all DA samples fall in the 2.5-9.5 μs range, i.e. they are up to ten times shorter than the 27 μs of the reference $x=0.015$ sample, whose Mn concentration is very similar to $x_{DA} \approx 0.013$. This demonstrates that our goal to tune independently static and dynamic magnetic properties of DMS materials is achieved by introducing digital alloys.

In order to give a detailed insight in the capabilities offered by digital alloys we compare the DA data points with the $\tau_{SLR}(x)$ dependence known for disordered alloys Cd$_{1-x}$Mn$_x$Te [4, 13, 14]. The latter is shown by the solid line in Fig. 3 (a). The data points for DA samples are positioned on this line according to their $\tau_{SLR}$ values, from which an effective Mn concentration $x'_{DA}$ can be assigned to each DA. The $x'_{DA}$ values range from 0.025 up to 0.039 and exceed significantly $x_{DA} \approx 0.013$.

Another instructive way to present the tuneability of the DA parameters is given in Fig. 3 (b). Here the correspondence diagram for the effective Mn concentrations determined by the static (i.e. the giant Zeeman splitting, the abscissa $x''_{DA}$) and the dynamic (i.e. the SLR time, ordinate $x'_{DA}$) magnetic properties in DMS is shown. For disordered alloys these effective concentrations are equal to the "real" Mn content, i.e. $x' \equiv x'' \equiv x$. As a result, the disordered alloys are described in the diagram by the straight solid bisecting line. The DA data points do not fall onto this line, confirming that the static and dynamic parameters of magnetization in the DMS digital alloys can be tuned separately.

Let us now discuss the physical reasons for providing the digital alloy flexibility. The origin lies in the different mechanisms responsible for the static magnetic properties, i.e. the giant Zeeman splitting of excitons measured here, and the SLR dynamics. The giant Zeeman splitting at relatively low Mn concentrations studied here is proportional to the number of Mn ions and to the overlap of the exciton wave function with a specific Mn ion. Both these quantities do not vary significantly in wide magnetic QWs with digital alloys. As demonstrated by Fig. 2 (a) the approach of using the average Mn content for predicting the value of the giant Zeeman splitting in case of Cd$_{0.95}$Mn$_{0.05}$Te/CdTe DA works very good.

In contrast, the SLR rate of the Mn ions is extremely sensitive to the probability to find another Mn ion at the nearest-neighbour or next-to-nearest-neighbour position, i.e. it is very sensitive to the magnetic ions clustering. The reason for that is that the spin-lattice relaxation of $Mn^{2+}$ ions in II-VI semiconductors is controlled by anisotropic Mn-Mn magnetic interactions in the Mn clusters [13, 14]. It is obvious from the scheme of Mn distribution in digital alloys (see Fig. 1), that the clustering probability is reduced in the 1×3DA structure due to the two-dimensional arrangement of the magnetic ions, but it increases significantly in the 3×9DA sample due to formation of clusters along the growth direction.

To conclude, we have demonstrated that the digital growth technique for DMS heterostructures gives access to controllable dynamics of the Mn spin system. We note here that a spatially inhomogeneous profile of Mn ions has been suggested very recently for control of SLR dynamics in heteromagnetic nanostructures [15]. The new possibilities offered by DMS digital alloys can be easily implemented to the heteromagnetic nanostructure concept.

**Acknowledgements.** We are thankful to A. A. Maksimov and A. V. Akimov for fruitful discussions. This work has been supported by the BMBF "nanoquit", DARPA QuIST and INTAS (Grant No. 03-51-5266).

**References**

[1] "*Semiconductor Spintronics and Quantum Computation*", ed. by D. D. Awschalom, D. Loss, and N. Samarth (Springer, Heidelberg 2002).

[2] T. Dietl, *Diluted magnetic semiconductors,* in: Handbook of semiconductors, Vol. 3b, ed. by S. Mahajan (North-Holland, Amsterdam 1994) p.1252.

[3] T. Dietl, P. Peyla, W. Grieshaber, and Y. Merle d'Aubigne, Phys. Rev. Lett. **74**, 474 (1995).

[4] M. K. Kneip, D. R. Yakovlev, M. Bayer, A. A. Maksimov, I. I. Tartakovskii, D. Keller, W. Ossau, L. W. Molenkamp, and A. Waag, cond-mat/0505485 19may2005; to be published in Phys. Rev. B.

[5] M. Kawabe, M. Kondo, N. Matsuuara, and Kenya Yamamoto, Jpn. J. Appl. Phys. **22**, L64 (1983).

[6] R. C. Miller, A. C. Gossard, D. A. Kleiman, and O. Munteanu, Phys. Rev. B **29**, 3740 (1983).

[7] A. C. Gossard, M. Sundaram, P. F. Hopkins, in: *Epitaxial Microstructures*, Ed. A. C. Gossard, in series: *Semiconductors and Semimetals*, Vol. 40, Academic Press, Boston 1994, p. 153.

[8] T. Wojtowicz, G. Karczewski, A. Zakrzewski, M. Kutrowski, E. Janik, E. Dynowska, K. Kopalko, S. Kret, J. Kossut and J. Y. Laval, Acta Phys. Pol. **A87** 165 (1995).

[9] S. A. Crooker, D. A. Tulchinsky, J. Levy, D. D. Awschalom, R. Garsia, and N. Samarth, Phys. Rev. Lett. **75**, 505 (1995).

[10] T. Wojtowicz, M. Kutrowski, G. Cywinski, G. Karczewski, E. Janik, E. Dynowska, J. Kossut, R. Fiederling, A. Pfeuffer-Jeschke and W. Ossau, J. Cryst. Growth **184/185** 936 (1998).

[11] W. J. Ossau and B. Kuhn-Heinrich, Physica B **184**, 442 (1993).

[12] W. Grieshaber, A. Haury, J. Cibert, Y. Merle d'Aubigne, A. Wasiela, and J. A. Gaj, Phys. Rev. B **53**, 4891 (1996).

[13] A. V. Scherbakov, A. V. Akimov, D. R. Yakovlev, W. Ossau, G. Landwehr, T. Wojtowicz, G. Karczewski, and J. Kossut, Phys. Rev. B **62**, R10641 (2000).

[14] D. Scalbert, Phys. Stat. Sol. (b) **193**, 189 (1996).

[15] A. V. Scherbakov, A. V. Akimov, D. R. Yakovlev, W. Ossau, L. Hansen, A. Waag, and L. W. Molenkamp, Appl. Phys. Lett. **86**, 162104 (2005).

**Figure Captions**

Fig. 1   Schematic diagram of the conduction and valence band profile and the Mn ion profile in a (Cd,Mn)Te digital alloy structures.

Fig. 2 (a) The Giant Zeeman shift of the photoluminescence line ($\sigma^+$ polarized) for three different DA samples (see text). The PL is excited by a *cw* semiconductor laser with the power density 0.22 W/cm$^2$. Lines show best fits to the data using Eqs. (1) and (2), to determine the average Mn content of the digital alloys; $S_{eff} = 2.5$ and $T_0 = 1$ K. In the inset PL spectra for the 1×3DA (solid line) and the 3×9DA (dashed) samples at *B*=0 and 10 T are given.

(b) Dynamical shift of the PL lines showing the cooling of the Mn spin system heated by pulsed laser excitation. Open symbols correspond to disordered alloys and closed symbols are for the digital alloys. SLR times were evaluated from mono-exponential fits.

Fig. 3 (a) SLR times versus Mn content *x* in disordered alloys (open circles are results of this paper and the solid line is taken from literature data). Closed circles are DA data assigned to the solid line due to their $\tau_{SLR}$ values. Their linking to the solid line allows evaluation of the effective Mn content $x'_{DA}$.

(b) Diagram linking the static (giant Zeeman splitting) and the dynamic (SLR time) magnetic characteristics of disordered (open circles and solid line) and digital (closed circles) alloys. In disordered alloys: $x' \equiv x'' \equiv x$.

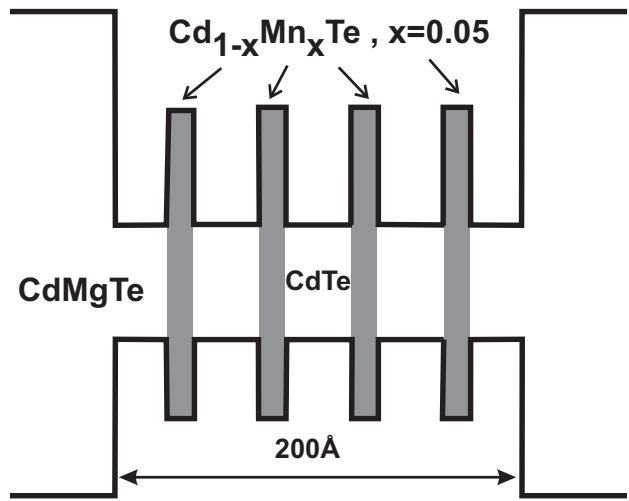
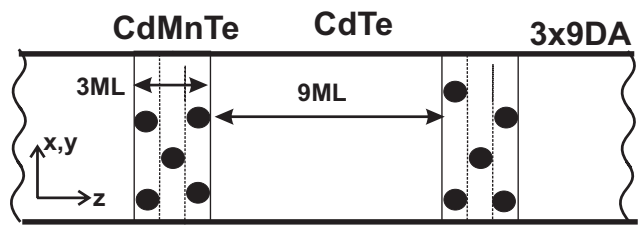
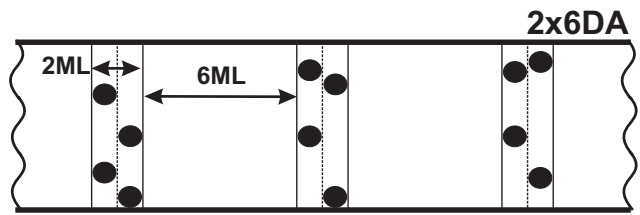
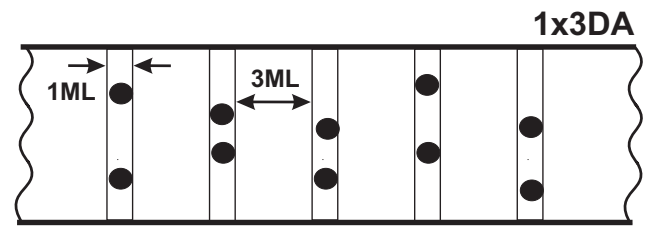

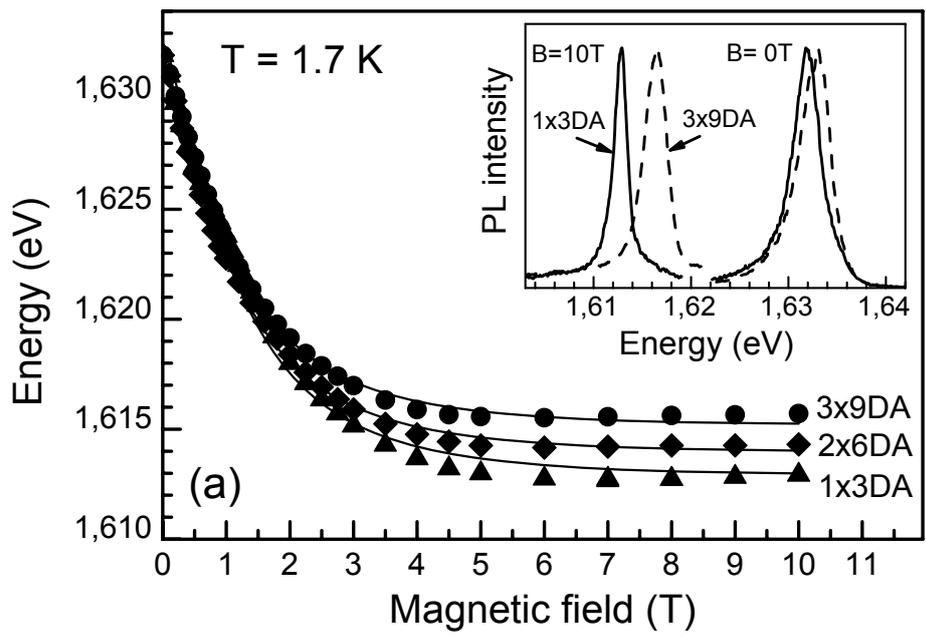
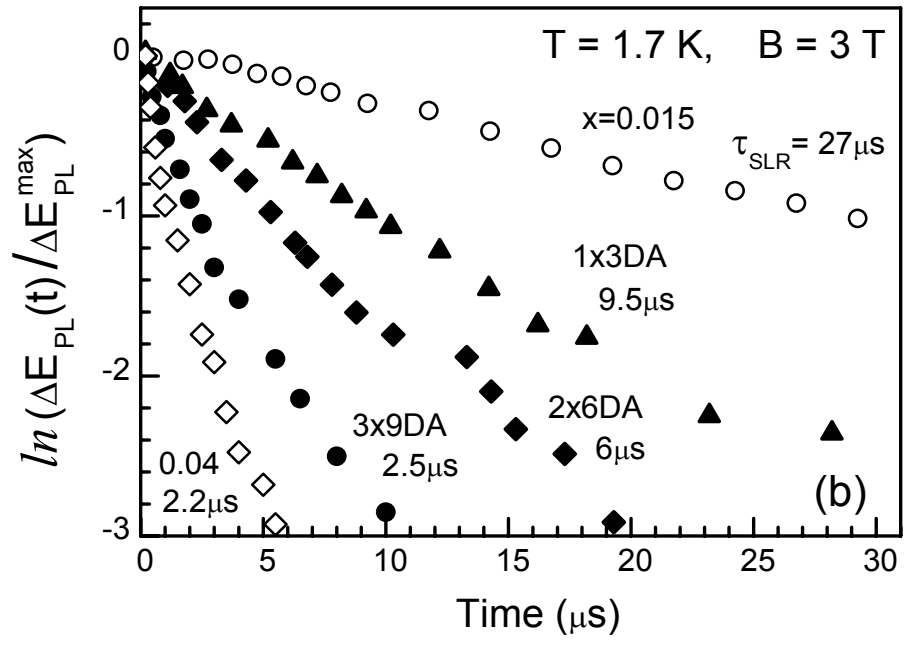

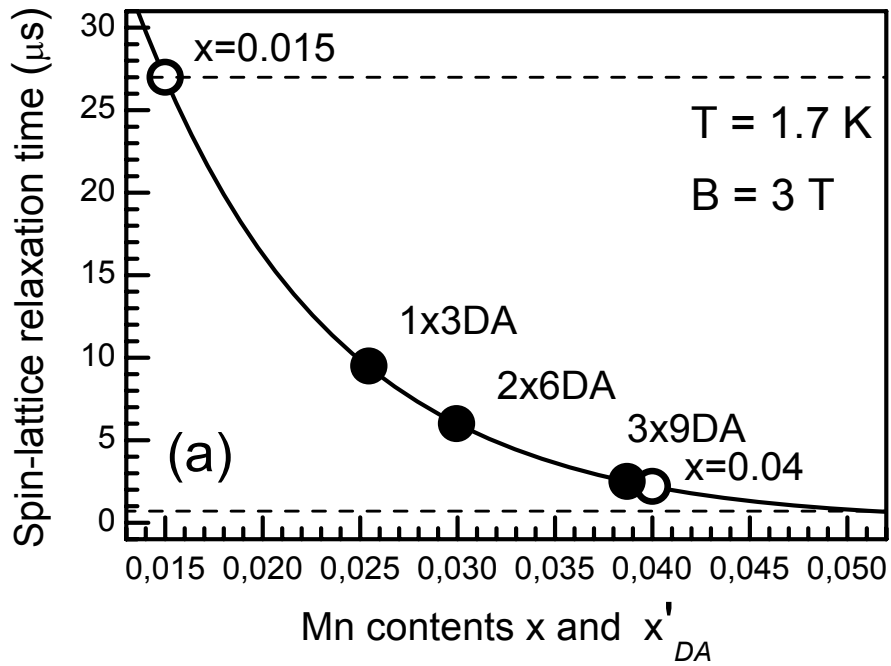

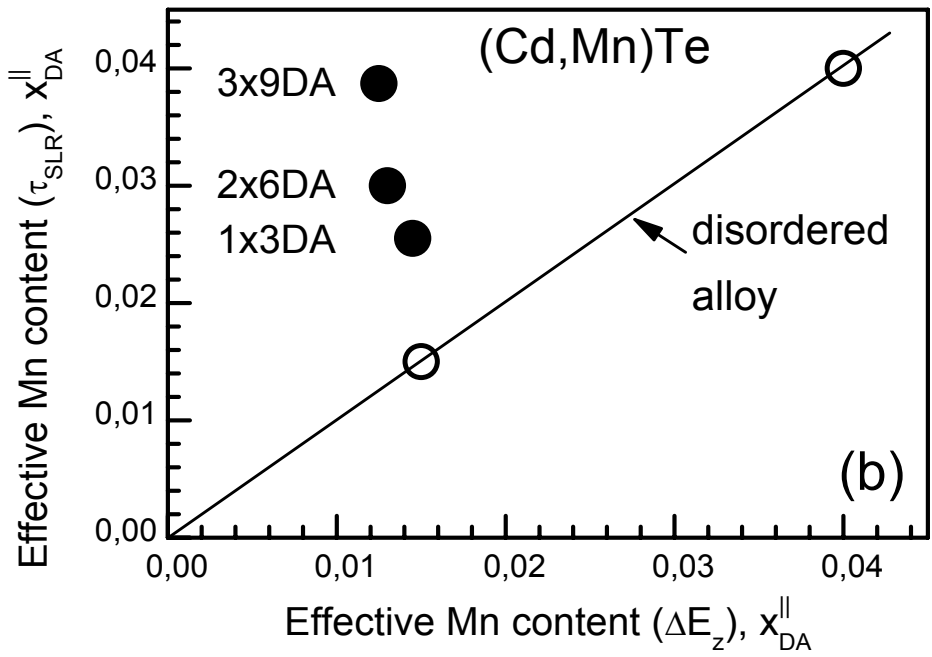